\documentclass[a4paper,fleqn,usenatbib]{mnras}

\usepackage[T1]{fontenc}
\usepackage{ae,aecompl}

\usepackage{graphicx}	% Including figure files
\usepackage{amsmath}	% Advanced maths commands
\usepackage{amssymb}	% Extra maths symbols

\renewcommand{\vec}[1]{\mbox{\boldmath $#1$}}
\title[Can superflares occur on the Sun?] % Short title 45 symbols maximum
{Dynamo model for grand maxima of solar activity: can superflares occur on the Sun?}
\author[L.\,L.~Kitchatinov \& S.\,V.~Olemskoy]
{L.\,L.~Kitchatinov$^{1,2}$\thanks{E-mail: kit@iszf.irk.ru}
and S.\,V.~Olemskoy$^{1}$
\\
$^{1}$Institute for Solar-Terrestrial Physics, Lermontov Str. 126A, Irkutsk 664033, Russia\\
$^{2}$Pulkovo Astronomical Observatory, St. Petersburg 196140, Russia
}

\date{Accepted XXX. Received YYY; in original form ZZZ}

\pubyear{2016}

\begin{document}
\label{firstpage}
\pagerange{\pageref{firstpage}--\pageref{lastpage}}
\maketitle

%%%%%%%%%%%%%%%%%%%%%%%%%%%%%%%%%%%%%%%%%%%%%%%%%%%%%%%%%%%%%%%%%%%%%%%%
\begin{abstract}
Recent data on superflares on Sun-like stars and radiocarbon data on solar activity in the past are both indicative of transient epochs of unusually high magnetic activity. We propose an explanation for the grand activity maxima in the framework of a solar dynamo model with fluctuating parameters. Solar-type dynamos are oscillatory because of the combination of the solar-type differential rotation with positive (in the Northern hemisphere) alpha-effect. An artificial reversal of the sign in the alpha-effect changes the dynamo to a steady regime with hundreds of times larger magnetic energy compared to the amplitude of the cyclic dynamo. Sufficiently large and durable fluctuations reversing the sign of the alpha-effect during the growth phase of a magnetic cycle can, therefore, cause a transient change to a steady dynamo with considerably increased magnetic energy. This qualitative scenario for grand activity maxima is supported by computations of the dynamo model with a fluctuating alpha-effect. The computed statistics of several thousand magnetic cycles gives examples of cycles with very high magnetic energy. Our preliminary estimations however suggest that the probability of solar superflares is extremely low.
\end{abstract}

% Select between one and six entries from the list of approved keywords.
% Don't make up new ones.
%%%%%%%%%%%%%%%%%%%%%%%%%%%%%%%%%%%%%%%%%%%%%%%%%%%%%%%%%%%%%%%%%%%%%%%
\section{Introduction}
%%%%%%%%%%%%%%%%%%%%%%%%%%%%%%%%%%%%%%%%%%%%%%%%%%%%%%%%%%%%%%%%%%%%%%%
The high-precision photometry of the {\sl Kepler} mission \citep{Kepler}, primarily aimed at detecting slight dimming in stellar luminosity caused by planetary transits, provided exciting new knowledge on stellar flares. Though only very powerful flares ($> 10^{33}$\,erg), exceeding the highest energy ($\sim 10^{32}$\,erg) of ever observed solar flares, could be detected in the photometric data, the number of stellar \lq superflares', reported in the pioneering works of \citet{Mea12} and \citet{Sea13}, exceeded the stellar flare statistics of the pre-{\sl Kepler} epoch considerably.

Many of the superflaring stars were solar-like G-dwarfs with long rotation periods $P_\mathrm{rot} > 10$\,days \citep{Sea13}. Among them, there were two stars very close to the Sun in their surface temperature and rotation rate \citep{Nea14}. This leads to the question of whether the superflares can occur on the Sun? Apart from its general interest, the question is of certain practical significance because the most powerful solar flares have caused damaging geomagnetic disturbances. \citet{Sea13} and \citet{Sea13PASJ} suggest a positive answer with an estimated frequency of one in about 800~yr for solar superflares of $\geq 10^{34}$\,erg (one in about 5000~yr for $\geq 10^{35}$\,erg).

It is remarkable that the distribution of superflare events among the {\sl Kepler} targets is highly non-uniform. \citet{Sea13} and \citet{Cea14} found that only about 0.3\% of observed stars have shown superflares during six quarters of observations. Among the flaring stars, however, there were about five superflares per star on average. Superflares are, therefore, inherent to durable special states of abnormally high magnetic activity, not sparse rare events that occur with equal probability on all solar-type stars. This is reminiscent of grand maxima of solar activity inferred by \citet{USK07,Uea14} from the content of radionuclides in natural archives.

The present paper proposes a dynamo model for the grand maxima with an attempt to estimate the possibility of solar superflares. \citet{Sea13PASJ} have already discussed this possibility from the standpoint of dynamo theory. They concluded that superflares would be possible in an exceptionally long solar cycle sufficiently durable for the differential rotation to wind a strong toroidal field with magnetic flux and energy sufficient for superflares. Such a hypothetical long cycle would be a certain infraction in solar cyclic activity. We suggest a dynamo-mechanism for such infractions.

It should be noted that currently there is no standard model for the solar dynamo. The absence of measurements of internal magnetic fields precludes a selection of the most realistic model among those hitherto proposed. However, the advection-dominated dynamos with Babcock--Leighton (BL) type alpha-effect for regeneration of a poloidal magnetic field originally proposed by \citet{CSD95} and \citet{D95} seem to provide the best agreement with the surface observations. The model used in this paper belongs to this type. It includes fluctuations of the alpha-effect.

Solar cycles observed differ in amplitude and duration. Dynamo theory explains this variability by fluctuations in dynamo parameters \citep[another possible explanation is dynamical chaos in nonlinear dynamos;][]{TWK95}. Previous dynamo models with fluctuating parameters were mainly aimed at explaining grand minima of solar activity similar to the famous Maunder minimum. The models generally reproduce grand minima statistics known from the radionuclide data. Publications by \citet{Mea08}, \citet{KC11}, \citet{CK12}, \citet{OK13} and \citet{HPN14} can be quoted as recent works among the extensive literature on this subject. Now we look in the \lq opposite direction' with an attempt to explain the grand maxima.

Our model does not involve any new mechanism or ingredient not used in dynamo models with fluctuating parameters before. Only the amplitude of fluctuations is of primary importance. The key idea is briefly as follows. Solar-type $\alpha\Omega$-dynamos are cyclic because the $\alpha$-parameter of the alpha-effect is positive (in the Northern hemisphere). In this case, the new poloidal field generated by the alpha-effect from the toroidal field is opposite to the old poloidal field, from which the toroidal field was wound by differential rotation, and the fields reverse periodically in successive cycles. Each new cycle, therefore, destroys fields generated in previous cycle and replaces them by new reversed fields. At least, current theory expects the solar dynamo to work in this inefficient way. If the $\alpha$-parameter were negative, the alpha-effect would amplify the pre-existing poloidal field. A steady dynamo with a much stronger magnetic field can be expected in this case (the expectation is confirmed by computations in the next section). If fluctuations of the alpha-effect are so large that their amplitude exceeds the mean value, the sign of the alpha-effect can occasionally reverse in course of a growth phase of a magnetic cycle. Then, a new poloidal field of the same sign as the old one will be generated and a long and powerful cycle will set in. This is our concept for the global maxima as infractions in the \lq normal' course of solar cyclic activity. The qualitative concept will be supported by a numerical model.

The alpha-effect of our model is a particular implementation of the BL mechanism for poloidal field generation. The mechanism is related to the finite average tilt of the bipolar sunspot groups relative to the local line of latitude \citep[cf., e.g.,][]{C10}. The mean tilt-angle is positive and corresponds to the positive alpha-effect (we follow the convention that the tilt-angle is positive when the spots leading in rotational motion are closer to the equator than the following spots and negative otherwise). The characteristic mean value of the tilt-angle ($\sim  5^\circ$) is, however, small compared to the width of its  observation-based distribution \citep[see fig.\,11 of][]{H96}. The distribution extends far into the region of negative tilt-angles. Also the number of spot groups simultaneously present on the Sun is not large. Large fluctuations leading to the reversals of sign of the BL alpha-effect required by the above scenario for grand solar maxima are therefore possible. Parameters of the BL mechanism can be estimated from sunspot data \citep{E04,Dea10,KO11AL}. Estimations of \citet{OCK13} show that the amplitude of fluctuations in the corresponding alpha-effect exceeds the mean value.
%%%%%%%%%%%%%%%%%%%%%%%%%%%%%%%%%%%%%%%%%%%%%%%%%%%%%%%%%%%%%%%%%%%%%%%
\section{Dynamo model}
%%%%%%%%%%%%%%%%%%%%%%%%%%%%%%%%%%%%%%%%%%%%%%%%%%%%%%%%%%%%%%%%%%%%%%%
The dynamo model of this paper is almost identical to that of \citet{OK13} (hereafter OK13). The only slight difference is in the formulation of the alpha-effect. The model is therefore described here briefly.
%%%%%%%%%%%%%%%%%%%%%%%%%%%%%%%%%%%%%%%%%%%%%%%%%%%%%%%%%%%%%%%%%%%%%%%
\subsection{Model equations}
%%%%%%%%%%%%%%%%%%%%%%%%%%%%%%%%%%%%%%%%%%%%%%%%%%%%%%%%%%%%%%%%%%%%%%%
Our mean-field model is two-dimensional. Velocity $\vec V$ of the large-scale flow
\begin{equation}
    \vec{V} = \vec{e}_\phi \Omega f(r,\theta)r\sin\theta   +
    \frac{1}{\rho}\vec{\nabla}\times\left(\vec{e}_\phi\frac{\psi(r,\theta)}{r\sin\theta}\right)
    \label{1}
\end{equation}
and the large-scale magnetic field $\vec B$
\begin{equation}
    {\vec B} = {\vec e}_\phi B(r,\theta) + {\vec\nabla}\times
    \left({\vec e}_\phi\frac{A(r,\theta)}{r\sin\theta}\right)
    \label{2}
\end{equation}
are assumed to be axisymmetric about the rotation axis. In these equations, the usual spherical coordinates ($r,\theta,\phi$) are used, ${\vec e}_\phi$ is the azimuthal unit vector, $\Omega$ is the mean angular velocity, $f(r,\theta)$ is the normalized frequency of differential rotation, $\psi$ is the stream function of the meridional flow, $B$ is the toroidal magnetic field, and $A$ is the poloidal field potential.
The second term on the right-hand side of Eq.\,(\ref{1}) describes the axisymmetric meridional flow. Isolines of the stream function $\psi$ in meridional planes are the stream-lines of the flow. Similarly, isolines of the potential $A$ of Eq.\,(\ref{2}) are the field lines of an axisymmetric poloidal field.

Dynamo equations are normalized to dimensionless units to reduce the number of free parameters. Then, the model controlling parameters combine in two dimensionless numbers: the dynamo number
\begin{equation}
    {\cal D} = \frac{\alpha_0 \Omega R_\odot^3}{\eta_0^2}\
    \label{3}
\end{equation}
and the magnetic Reynolds number for the meridional flow
\begin{equation}
    R_\mathrm{m} = \frac{V_0 R_\odot}{\eta_0} ,
    \label{4}
\end{equation}
where $\alpha_0$ and $\eta_0$ are the characteristic values of the alpha-effect and the eddy magnetic diffusivity respectively, $R_\odot$ is the solar radius, and $V_0$ is the amplitude of the meridional flow velocity. Both key numbers enter the normalized dynamo equation for the toroidal field,
\begin{eqnarray}
    \frac{\partial B}{\partial t} &=&
    \frac{\cal D}{x} \left(\frac{\partial f}{\partial x}\frac{\partial
    A}{\partial\theta} - \frac{\partial f}{\partial\theta}
    \frac{\partial A}{\partial x}\right)
    \nonumber \\
    &+& \frac{R_\mathrm{m}}{x^2\rho}\frac{\partial}{\partial\theta}
    \left(\frac{B}{\sin\theta}
    \frac{\partial\psi}{\partial x}\right)
    - \frac{R_\mathrm{m}}{x\sin\theta}\frac{\partial}{\partial x}
    \left(\frac{B}{\rho x}\frac{\partial\psi}{\partial\theta}
    \right)
    \nonumber \\
    &+&  \frac{\eta}{x^2}\frac{\partial}{\partial\theta}\left(
    \frac{1}{\sin\theta}\frac{\partial(\sin\theta
    B)}{\partial\theta}\right)
    \nonumber \\
    &+& \frac{1}{x}\frac{\partial}{\partial
    x}\left(\sqrt{\eta}\ \frac{\partial(\sqrt{\eta}\ xB)}
    {\partial x}\right) ,
    \label{5}
\end{eqnarray}
where the same notations are kept for normalized variables as used before for their dimensional counterparts, except for the fractional radius $x = r/R_\odot$.  In Eq.\,(\ref{5}), $\eta$ is the eddy diffusivity $\eta_{_\mathrm{T}}$ normalized to its characteristic value $\eta_0$: $\eta(x) = \eta_{_\mathrm{T}}/\eta_0$ (see OK13 for the details of the normalization procedure). The dynamo equation accounts for the diamagnetic pumping with the effective velocity
\begin{equation}
    \vec{U}_\mathrm{dia} = - \frac{1}{2}\vec{\nabla}\eta_{_\mathrm{T}}
    \label{6}
\end{equation}
\citep[][chapter 9.5]{KR80}. The pumping effect is important for our dynamo model. Due to the pumping, the distributed model realises an essentially \lq interface dynamo' near the base of the convection zone \citep{KO12}.

The poloidal field equation,
\begin{eqnarray}
    \frac{\partial A}{\partial t} &=&
    x \cos\theta\sin^3\theta \left(1 + \sigma s(t)\right)
    \int\limits_{x_\mathrm{i}}^x
    \hat\alpha (x,x') B(x',\theta)\ \mathrm{d} x'
    \nonumber \\
    &+& \frac{R_\mathrm{m}}{\rho x^2 \sin\theta}
    \left(\frac{\partial\psi}{\partial x}
    \frac{\partial A}{\partial\theta} -
    \frac{\partial\psi}{\partial\theta}
    \frac{\partial A}{\partial x}\right)
    \nonumber \\
    &+& \frac{\eta}{x^2}\sin\theta\frac{\partial}{\partial\theta}
    \left(\frac{1}{\sin\theta}\frac{\partial
    A}{\partial\theta}\right) + \sqrt{\eta}\frac{\partial}{\partial
    x} \left(\sqrt{\eta}\frac{\partial A}{\partial x}\right)  ,
    \label{7}
\end{eqnarray}
accounts for random fluctuations in the field generation term. The first term on the right-hand side represents the non-local alpha-effect of BL type. $x_i$ in the integration term is the radius of the inner boundary. The integration proceeds up to the upper limit $x$, which qualitatively reflects the fact that the generation of poloidal fields at some point $x$ is contributed to by the rise of magnetic loops from deeper layers and that the rise is almost vertical \citep{DC93}. The term $\sigma s$ accounts for the random fluctuations in the generation process. $\sigma$ is the relative amplitude of the fluctuations and $s(t)$ is a random function of time of the order of 1. The only difference from OK13 is that $s$ is now a function of time only. OK13 studied the north--south asymmetry of the dynamo process induced by the dependence of fluctuations on latitude. Hemispheric asymmetry is not essential for the present paper and we omit the random dependence of fluctuations on latitude to speed-up the computations. The computations have to cover a long time to accumulate statistics of many dynamo-cycles. The computations of this paper were performed with the relative amplitude of fluctuations $\sigma = 2.7$ estimated from sunspot statistics \citep{OCK13}.

The random process is modelled by solving numerically the equation system,
\begin{eqnarray}
    \frac{\partial s}{\partial t} &=& -\frac{s}{\tau} + \frac{s_1}{\tau} ,
    \nonumber \\
    \frac{\partial s_1}{\partial t} &=& -\frac{s_1}{\tau}
    + \frac{2\hat{g}}{\sqrt{\tau \Delta t}} ,
    \label{8}
\end{eqnarray}
in line with the dynamo equations. In Eqs\,(\ref{8}), $\tau$ is the characteristic time of variation of $s$ (the correlation time), $\Delta t$ is the time-step of the numerical model, and $\hat{g}$ is the random number with Gaussian distribution, zero mean, and rms value equal to 1. The random number was renewed each time-step independently of its former value. For the case of a short time-step, $\Delta t \ll \tau$, equations (\ref{8}) realize a smoothly varying random function of time with the correlation
\begin{equation}
    \langle s(t_0 + t)s(t_0)\rangle = \left(\frac{\mid t \mid }{\tau} + 1\right)
    \mathrm{exp}\left(-\mid t\mid /\tau \right),
    \label{9}
\end{equation}
where the angular brackets signify averaging over time $t_0$. OK13 confirmed by computations that a numerical solution of Eqs\,(\ref{8}) reproduces the analytical correlation function (\ref{9}).

The inner boundary is placed at $x_\mathrm{i} = 0.7$ slightly below the helioseismologically detected position of the base of the convection zone at $x_b = 0.713$ \citep{BA97}. The conditions for the interface with superconductor
\begin{equation}
    \frac{\partial\left(\sqrt{\eta}xB\right)}{\partial x} = 0,\ \ \
    A = 0 \ \ \ \mathrm{at}\ x = x_\mathrm{i}
    \label{10}
\end{equation}
were imposed at this boundary. Pseudo-vacuum conditions
\begin{equation}
    B = 0,\ \ \ \frac{\partial A}{\partial x} = 0\ \ \ \mathrm{at}\ x = 1
    \label{11}
\end{equation}
were applied at the top.
%%%%%%%%%%%%%%%%%%%%%%%%%%%%%%%%%%%%%%%%%%%%%%%%%%%%%%%%%%%%%%%%%%%%%%%
\subsection{Model design}
\begin{figure}
	\includegraphics[width=\columnwidth]{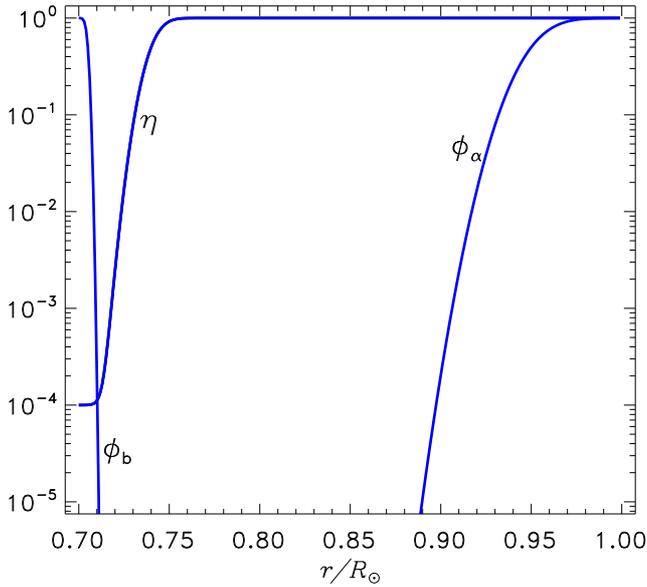}
    \caption{Profiles of the normalized diffusivity $\eta$ and kernel functions
        (\ref{13}) of the non-local alpha-effect of the dynamo equation (\ref{7}) used in our model.}
    \label{f1}
\end{figure}

%%%%%%%%%%%%%%%%%%%%%%%%%%%%%%%%%%%%%%%%%%%%%%%%%%%%%%%%%%%%%%%%%%%%%%%
Eddy diffusion varies mildly in the bulk of the convection zone but decreases sharply with depth near its base. This is reflected in the diffusivity profile
\begin{equation}
    \eta (x) = \eta_\mathrm{in} + \frac{1}{2}(1 - \eta_\mathrm{in})
    \left( 1 + \mathrm{erf}\left(\frac{x -
    x_\eta}{h_\eta}\right)\right) ,
    \label{12}
\end{equation}
where $\eta_\mathrm{in}$ is the ratio of the overshoot region diffusivity to its value in the convection zone. The diffusivity profile for the values $\eta_\mathrm{in} = 10^{-4}$, $x_\eta = 0.74$ and $h_\eta = 0.01$ used in our computations is shown in Fig.\,\ref{f1}. The same figure shows the kernel functions of the non-local alpha-effect of the dynamo equation (\ref{7}):
\begin{eqnarray}
    \hat\alpha (x,x') &=& \frac{\phi_\mathrm{b}(x')\phi_\alpha (x)} {1 + B^2(x',\theta)}
    ,
    \nonumber \\
    \phi_\mathrm{b}(x') &=& \frac{1}{2}\left( 1 -
    \mathrm{erf}\left( (x' - x_\mathrm{b})/h_\mathrm{b}\right)\right) ,
    \nonumber \\
    \phi_\alpha (x) &=& \frac{1}{2}\left( 1 +
    \mathrm{erf}\left( (x - x_\alpha)/h_\alpha\right)\right).
    \label{13}
\end{eqnarray}
The function $\phi_\mathrm{b}(x')$ of this equation defines the region near the bottom boundary whose toroidal fields produce the alpha-effect near the top. The function $\phi_\alpha(x)$ defines the near-surface region where the alpha-effect generates poloidal fields. The values $h_\alpha = 0.02$ and $h_\mathrm{b} = 0.002$ of the parameters of Eq.\,(\ref{13}) were used in the computations. The prescriptions $x_\alpha = 1 - 2.5h_\alpha$ and $x_\mathrm{b} = x_\mathrm{i} + 2.5h_\mathrm{b}$ ensure smoothness of the kernel functions (Fig.\,\ref{f1}).

The diffusivity of Fig.\,\ref{f1} decreases strongly near the bottom boundary. The small near-bottom diffusivity and the downward diamagnetic pumping (\ref{6}) result in fine structure of the magnetic field near the bottom. The numerical grid of our model is therefore not uniform in radius. The grid spacing is proportional to the square root of the local diffusivity value, $\Delta x \sim \sqrt{\eta}$.
With this choice, the numerical stability criterion for the fully explicit time-stepping of our model is uniform with radius. The grid-point method of the model is second-order accurate in space and first-order accurate in time. The time step is, however, proportional to the second order of the radial grid spacing $\Delta t = \varepsilon \Delta x^2\eta^{-1}$ ($\varepsilon = 0.3$ ensured numerical stability). Computations were performed with a grid of $201\times201$ points in latitude and radius. The grid is uniform in latitude. No boundary conditions restricting the field parity were imposed at the equator.

\begin{figure}
	\includegraphics[width=\columnwidth]{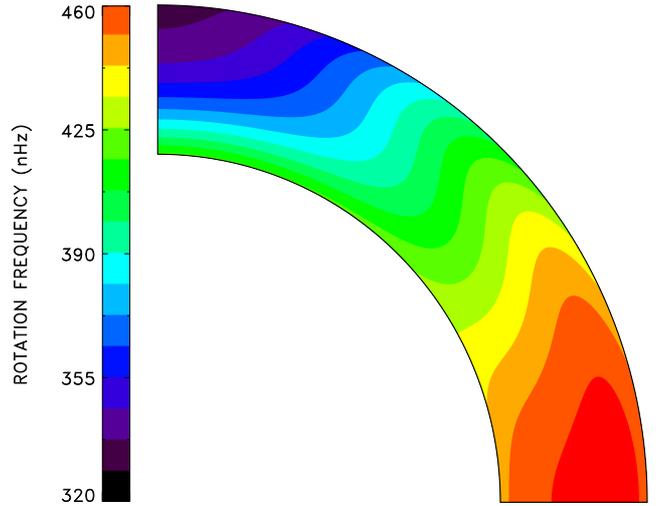}
    \caption{Iso-contours of rotation frequency for the differential
        rotation of our dynamo model.}
    \label{f2}
\end{figure}

The differential rotation was specified with the approximation of \citet{BKS00} for the helioseismological rotation law:
\begin{equation}
    f(x,\theta) = \frac{1}{461}\sum\limits_{m=0}^{2}
    \cos\left( 2m\left(\frac{\pi}{2}
    - \theta\right)\right) \sum\limits_{n=0}^{4} C_{nm}x^n ,
    \label{14}
\end{equation}
where $f$ is the normalized rotation frequency of Eq.\,(\ref{1}),
numerical values of the coefficients $C_{nm}$ are given in \citeauthor{BKS00}. The rotation law is shown in Fig.\,\ref{f2}.

The single-cell meridional circulation with poleward flow on the top and the return flow towards the equator near the bottom was specified with the following prescription for the stream-function:
\begin{eqnarray}
       \psi &=&- \cos\theta\ \sin^2\theta\ \Phi(x) ,
    \nonumber  \\
    \Phi(x) &=& \left\{\begin{array}{ll}
            \frac{1}{1 - x_\mathrm{s}} \int\limits_x^1 \rho
            (x')\eta^p(x')(x'-x_\mathrm{s}) \rm{d} x' & \mbox{for $x
            \geq x_\mathrm{s}$} \\
            C \int\limits_{x_\mathrm{i}}^x \rho(x') \eta^p(x')
            (x_\mathrm{s} - x') \rm{d} x' & \mbox{for $x
            \leq x_\mathrm{s} .$}
            \end{array}
            \right.
            \label{15}
\end{eqnarray}
where $x_\mathrm{s}$ is the fractional radius of the stagnation point where the meridional velocity changes sign and $C$ is a constant factor whose value is adjusted to ensure continuity of the stream-function at the stagnation radius. The adiabatic profile,
\begin{equation}
    \rho(x) = \left( 1 + C_\rho\left(\frac{1}{x}
    -1\right)\right)^{3/2} ,
    \label{16}
\end{equation}
with $C_\rho = 10^3$ was used for the normalized density. Computations were performed with $p = 1/3$ and $x_\mathrm{s} = 0.77$. With such a deep stagnation point, the velocity of the near-bottom flow is only slightly smaller compared to the surface flow (see fig.\,3 of OK13). This flow structure is similar to the latest helioseismic data of \citet{RA15} and to the mean-field model of global solar circulation \citep{KO11}.

The magnetic Reynolds number (\ref{4}) in the computations to follow is $R_\mathrm{m} = 10$. This implies a background diffusivity value $\eta_0 = 10^9$\,m$^2$s$^{-1}$, which is favoured by mixing-length estimations. Conversions between dimensionless and physical time-scales will be performed with this diffusivity value. In particular, the dimensionless rotation period is estimated to be
\begin{equation}
    p_\mathrm{rot} = P_\mathrm{rot}\eta_0 R_\odot^{-2} = 4.53\times 10^{-3} ,
    \label{17}
\end{equation}
where $P_\mathrm{rot} = 25.4$\,day is the sidereal period of solar rotation. The lifetime of sunspot groups is of the same  order as the rotation period \citep{S03}. The correlation time of the fluctuations $\tau = 5\times 10^{-3}$ was taken close to the normalized rotation period (\ref{17}).
%%%%%%%%%%%%%%%%%%%%%%%%%%%%%%%%%%%%%%%%%%%%%%%%%%%%%%%%%%%%%%%%%%%%%%%
\subsection{Properties of the model without fluctuations}
%%%%%%%%%%%%%%%%%%%%%%%%%%%%%%%%%%%%%%%%%%%%%%%%%%%%%%%%%%%%%%%%%%%%%%%
The results of the next section can be interpreted in terms of the model properties for the case of vanishing fluctuations. We therefore consider first the case of $\sigma = 0$ in Eq.\,(\ref{7}).

The critical value of dynamo number (\ref{3}) for the onset of dynamo-instability is ${\cal D}_\mathrm{c}^\mathrm{d} = 3.01\times 10^4$. The upper index \lq d' means that the threshold dynamo number corresponds to the  dipolar (equator-antisymmetric) magnetic field. The threshold value for the generation of quadrupolar fields, ${\cal D}_\mathrm{c}^\mathrm{q} = 3.79\times 10^4$, is higher.
Accordingly, the magnetic field converges to dipolar parity irrespectively of its initial equatorial symmetry.
The preference for dipolar fields is a common feature of solar dynamo models with a strong decrease of diffusivity towards the inner boundary \citep{CNC04,JCC07,HY10}.
The initial field in all runs of this paper combined the toroidal field of dipolar parity, $B_0 = 0.1(x - x_\mathrm{i})^2(1-x)\sin(2\theta)$, and zero poloidal field. In several diffusive times, the field dynamics approach a regime independent of the field initial amplitude or spatial distribution. The memory time of the solar dynamo is comparable to the length of an activity cycle \citep{Yea08}. The results to follow correspond to these asymptotic regimes independent of the initial state.

The solar dynamo is probably only slightly supercritical \citep[see discussion in][]{KKB15}. All the computations to follow are therefore performed with only 5 per cent supercritical ${\cal D} = 3.16\times 10^4$.

\begin{figure}
	\includegraphics[width=\columnwidth]{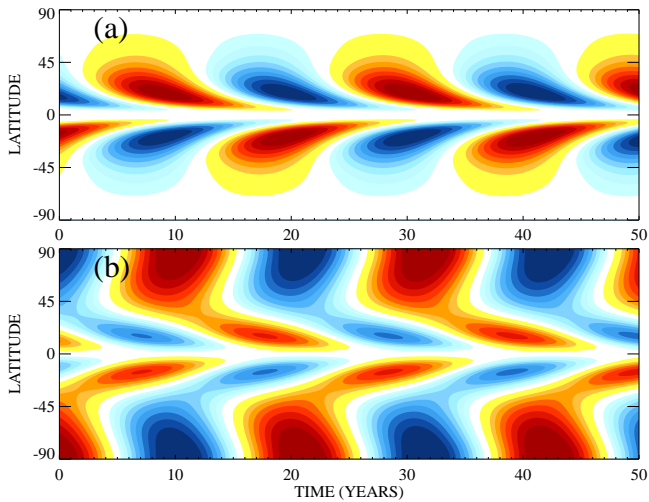}
    \caption{Time--latitude diagrams for the near-bottom toroidal flux
        density (\ref{18}) (a) and surface radial field (b) for the model
        without fluctuations.}
    \label{f3}
\end{figure}

Figure~\ref{f3} shows the time--latitude diagrams of magnetic oscillations for the case of vanishing fluctuations. The BL mechanism is related to sunspot activity. The butterfly diagram of Fig.\,\ref{f3} therefore shows the same near-bottom flux density per unit length in latitude,
\begin{equation}
    {\cal B} = \sin\theta \int\limits_{x_\mathrm{i}}^1
    \phi_\mathrm{b}(x) B(x)\ \mathrm{d}x ,
    \label{18}
\end{equation}
as present in the alpha-effect of equations (\ref{7}) and (\ref{13}). The factor $\sin\theta$ is included in this equation because of the assumption that sunspot production is proportional to the length of the toroidal flux tube.

The field patterns of Fig.\,\ref{f1} roughly reproduce the observed dynamics of solar large-scale magnetic fields. The toroidal fields near the bottom boundary are about one thousand times larger than the surface polar fields. The amplitude of the normalized magnetic energy,
\begin{equation}
    W = \frac{1}{4}\int\limits_0^\pi\int\limits_{x_\mathrm{i}}^1
    x^2 \sin\theta B^2 \mathrm{d}\theta\mathrm{d}x ,
    \label{19}
\end{equation}
for the oscillatory dynamo was $W_0 = 7.4 \times 10^{-5}$ (it should be multiplied by about $10^{40}$ for conversion into dimensional cgs units). We shall use the fractional magnetic energy $\hat{W} = W/W_0$ normalized to this \lq representative' value for comparison with other dynamo regimes.

\begin{figure}
	\includegraphics[width=\columnwidth]{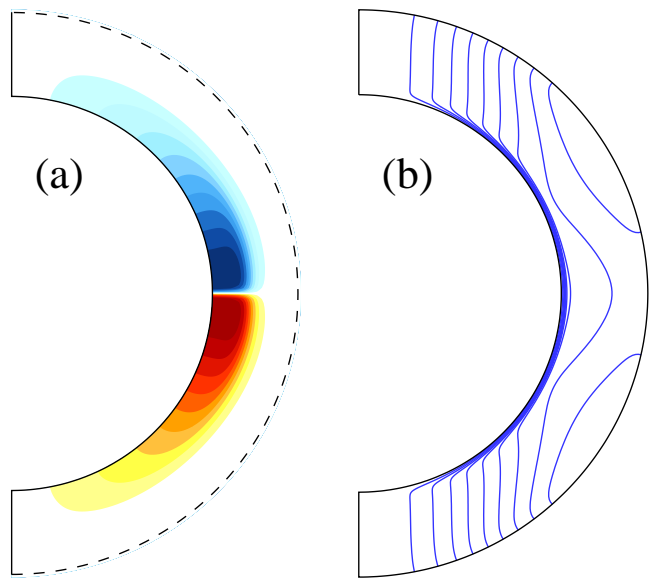}
    \caption{Toroidal field levels (a) and the poloidal field lines (b)
        for a steady dynamo with negative alpha-effect, ${\cal D} = -3.16\times 10^4$. The radial scale in part (a) is augmented by placing the upper (dashed) boundary at $x = 0.74$. The poloidal field direction corresponds to the clock-wise circulation. Accordingly, the toroidal field is negative in the Northern and positive in the Southern hemisphere.}
    \label{f4}
\end{figure}

If the sign of the alpha-effect is reversed, the dynamo changes from oscillatory to steady (by sign reversal, we mean the change of sign in the first term of the right-hand side of Eq.\,(\ref{7}) or in dynamo number (\ref{3})). The steady field pattern for $D = -3.16\times 10^4$ is shown in Fig.\,\ref{f4}. The upper (dashed) boundary in the toroidal field pattern of this figure is at the radius $x = 0.74$. The toroidal field is concentrated near the bottom boundary due to the diamagnetic pumping of Eq.\,(\ref{6}). The poloidal field is concentrated at the bottom as well but penetrates to the surface because the radial field is not affected by the radial pumping. Our distributed model, therefore, realizes an essentially interface dynamo due to allowance for the pumping effect \citep{KO12}. The (fractional) magnetic energy $\hat{W} = 440$ for the steady dynamo with negative alpha-effect is large compared to the oscillatory dynamo of Fig.\,\ref{f3} with a positive alpha. We can therefore expect a strong increase in magnetic energy from infractions in the dynamo mechanism caused by stochastic reversals of the sign of the alpha-effect.
\begin{figure*}
	\includegraphics[width=18 truecm]{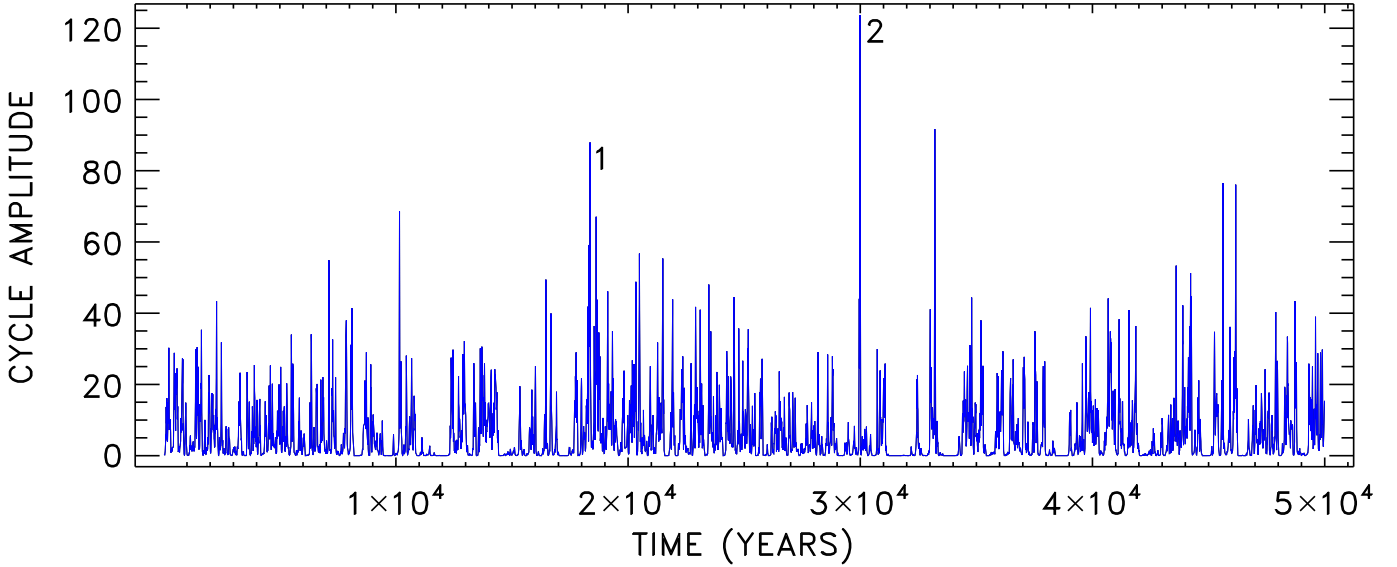}
    \caption{Amplitudes $\hat{W}$ of the dynamo-cycles for the
        computation covering about 50000 years. Positions of neighbouring cycles on the time-amplitude diagram are connected by straight lines to visualize these positions. More details on the high cycles labelled 1 and 2 in this figure are shown in Figs 6 and 7.
        }
    \label{f5}
\end{figure*}
%%%%%%%%%%%%%%%%%%%%%%%%%%%%%%%%%%%%%%%%%%%%%%%%%%%%%%%%%%%%%%%%%%%%%%%
\section{Results}
%%%%%%%%%%%%%%%%%%%%%%%%%%%%%%%%%%%%%%%%%%%%%%%%%%%%%%%%%%%%%%%%%%%%%%%
Fluctuations in the alpha-effect change the dynamo from the strictly periodic regime of Fig.\,\ref{f3} to cycles with variable amplitudes and durations. Figure~\ref{f5} shows the amplitudes of about four thousand computed cycles in terms of their fractional magnetic energy $\hat{W}$. We use the cycle amplitude $W_0$ for zero fluctuations as the representative value to which the amplitudes of irregular cycles for finite fluctuations are normalized. The mean normalized amplitude $\langle \hat{W}\rangle = 5.9$ is, however, not close to one (the angular brackets mean averaging over the ensemble of computed cycles). This relatively large mean amplitude results from the asymmetry between high- and low-amplitude cycles caused by the positive definiteness of magnetic energy: two cycles of ten times smaller and ten times larger amplitude compared to $W_0$ have on average an amplitude of about $5W_0$. The \lq characteristic amplitude' $\mathrm{exp}\langle \ln( W )\rangle =1.007 W_0$ of the computed cycles is, however, close to $W_0$.

\begin{figure}
	\includegraphics[width=\columnwidth]{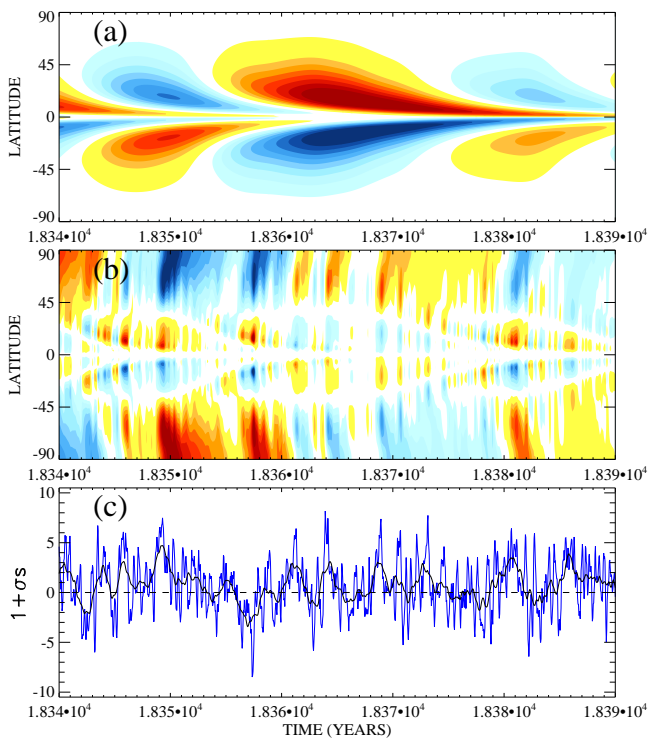}
    \caption{Evolution of the magnetic field and fluctuating alpha-effect
        near the high cycle marked by 1 in Fig.\,\ref{f5}. (a) The butterfly diagram of the near-bottom toroidal flux density (\ref{18}). (b) The time--latitude pattern of the surface poloidal field. (c) The $1+\sigma s$ -- coefficient of the alpha-effect in Eq.\,(\ref{7}). The thick line shows the annual running mean of this coefficient.
        }
    \label{f6}
\end{figure}

\begin{figure}
	\includegraphics[width=\columnwidth]{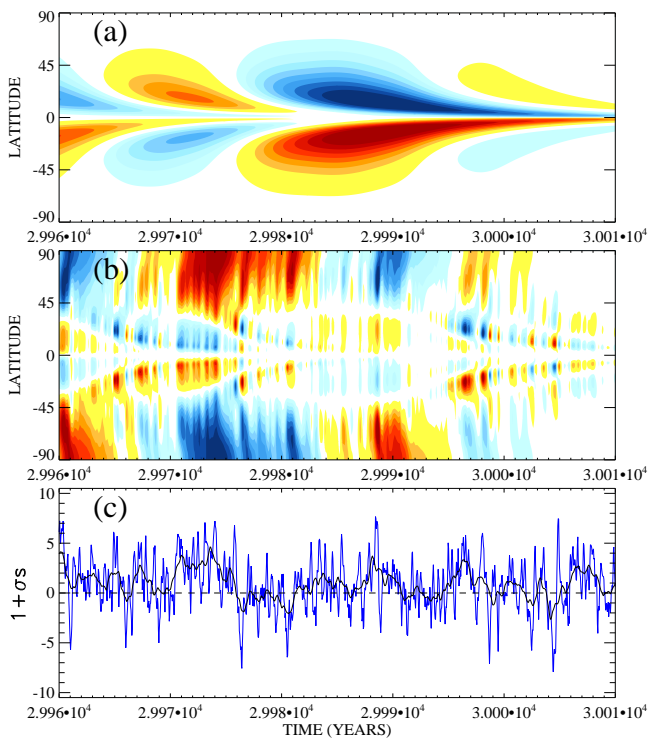}
    \caption{The same as in Fig.\,\ref{f6} but for the high cycle marked by 2 in Fig.\,\ref{f5}.}
    \label{f7}
\end{figure}

Fig.~\ref{f5} shows that cycles of much higher magnetic energy compared to $W_0$ are sometimes met in the very long computation. These high cycles are caused by the infractions in the dynamo process discussed in the Introduction. Figs \ref{f6} and \ref{f7} unfold the dynamics of magnetic fields and fluctuating alpha-effect around the high cycles labelled 1 and 2 in Fig.\,\ref{f5}. In both cases, there were durable reversals of the sign of the alpha-effect in the growth phase of the high cycles. By the \lq growth phase',we mean the epoch of toroidal field amplification. Normally, the poloidal field declines in the growth phase. The decline prevents winding of \lq too strong' toroidal fields by differential rotation. An occasional reversal of the alpha-effect in the growth phase amplifies the poloidal field so that an abnormally strong and durable cycle sets in.

The high cycles of Figs \ref{f6} and \ref{f7} obey this scenario. Durable reversals of the alpha-effect are typical of the high cycles of our model. We therefore propose infractions in the cyclic dynamo caused by statistical reversals of the alpha-effect as the mechanism for grand activity maxima.

The poloidal field diagrams of Figs \ref{f6} and \ref{f7} are relatively fine-structured. The structuring is caused by the fluctuations in the alpha-effect of the poloidal field generation. The fine structure is smoothed out while the field is transported to the base of the convection zone where the toroidal field is wound and the toroidal field patterns are smooth.

Unlike the grand minima, which normally comprise several or many magnetic cycles, high cycles are usually alone. Grand maxima of several neighbouring cycles with $W > 10W_0$ occurred seldom in our computations.

\begin{figure}
	\includegraphics[width=\columnwidth]{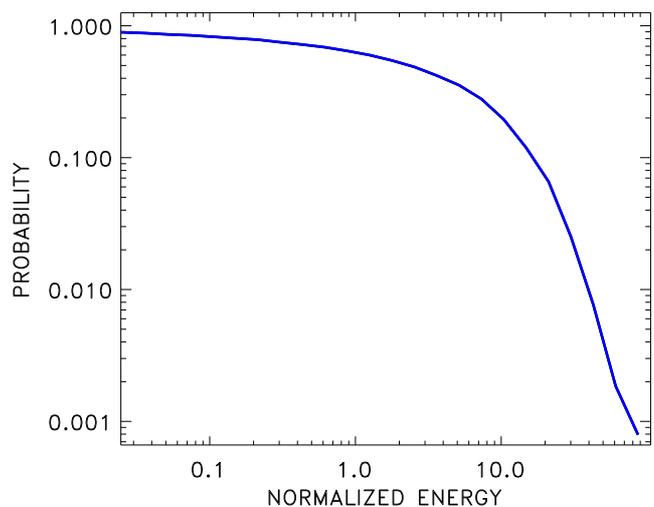}
    \caption{Probability for the amplitude of a magnetic cycle to exceed
        a given value of fractional magnetic energy. }
    \label{f8}
\end{figure}

The probability of the amplitude of a magnetic cycle exceeding a given value $\hat{W}$ of fractional magnetic energy is shown in Fig.\,\ref{f8} as a function of $\hat{W}$. This plot can be understood also as a fraction of computed cycles with amplitudes higher than $\hat{W}$. The probability of a dynamo cycle exceeding $10W_0$ in amplitude is $P(>10) = 0.204$ (Fig.\,\ref{f8}). About 2.5 per cent of computed cycles had amplitudes in excess of $30W_0$ and only one cycle (marked 2 in Fig.\,\ref{f5}) had an amplitude in excess of $100W_0$. If we assume linear proportionality between the maximum flare energy and the energy of dynamo-generated magnetic fields, then solar cycles producing superflares of $10^{34}$\,erg should be extremely seldom, that is, to be met only once in several thousand activity cycles.
%%%%%%%%%%%%%%%%%%%%%%%%%%%%%%%%%%%%%%%%%%%%%%%%%%%%%%%%%%%%%%%%%%%%%%%
\section{Conclusions}
%%%%%%%%%%%%%%%%%%%%%%%%%%%%%%%%%%%%%%%%%%%%%%%%%%%%%%%%%%%%%%%%%%%%%%%
We propose statistical reversals of the alpha-effect on the growth phase of solar or stellar activity cycles as a mechanism for the emergence of cycles with exceptionally high magnetic energy. The sign reversals cause infractions in the cyclic dynamo in the form of transient changes to a steady dynamo. The steady dynamo with reversed sign of the alpha-effect (Fig.\,\ref{f4}) has about 400 times higher magnetic energy compared to the amplitude of a cyclic dynamo. Even a short-term change to a steady regime, therefore, results in a significant increase in magnetic energy.

This scenario for the grand maxima of stellar activity is in some sense similar to the model of \citet{GRE05} for geomagnetic reversals. The difference, however, is that the \lq background' geodynamo of \citeauthor{GRE05} was steady. Random fluctuations in their model occasionally changed the alpha-effect to a state where a transient oscillatory dynamo sets in. Magnetic field reversals occurred if the transients were sufficiently durable. The reversals were accompanied by a significant reduction in magnetic energy. In our case, transient changes of the oscillatory dynamo to a steady regime are accompanied by an increase in magnetic energy. Geomagnetic reversals and grand maxima of stellar activity may be of similar origin.

Our model computations support the above scenario for grand activity maxima. Two examples of Figs \ref{f6} and \ref{f7} show sign reversals of the alpha-effect on the growth phase of high magnetic cycles. The supposed change of the dynamo to a transient steady regime explains the long durations of the high cycles. The longevity of high cycles agrees with the estimations of \citet{Sea13PASJ} showing that only durable cycles can produce sufficiently strong toroidal fields for superflares. If we assume that the maximum possible energy of flares is proportional to the energy of the internal stellar magnetic field and the maximum flare energy for a typical solar cycle is about $10^{32}$\,erg, then the simulations of this paper suggest that superflares of $\geq 10^{34}$\,erg should be extremely rare and practically impossible on the Sun. It seems, however, premature to claim such a conclusion.
Conversion of the energy of dynamo-generated fields into the energy released in flares is purely understood and represents an essentially open problem which is in the core of the relation between dynamo and flaring activity.
\citet{Cea14} associated spikes of Ohmic dissipation in their dynamo model with flaring events. It remains, however,  unclear how the energy of magnetic fibrils of which the large-scale dynamo-field probably consists \citep{P84} depends on the energy of the parent dynamo-field.

The unknown relation between the energies of flares and internal magnetic field of a star may not be linear. Also, our model includes purely known parameters.  The amplitude of the fluctuations was estimated from sunspot data \citep{OCK13}. The correlation time of the fluctuations, however, is difficult to estimate. It can be expected that high cycles will be more frequent in dynamo simulations with a longer correlation time. Simulations of magnetic cycle statistics demand long computations. We do not attempt to explore the parametric space already in this publication introducing the grand maxima model.

If the above scenario for grand activity maxima is correct, it may have a certain prognostic significance. The Babcock-Leighton mechanism is related to observable properties of sunspots \citep{Dea10,OCK13}. An inversion of the BL-type alpha-effect implies violation of Joy's law. Multiple violation of this law at the beginning of an activity cycle can be a predictor of the high amplitude and long duration of the coming cycle. A short-term violation of Joy's law has been actually noticed in the current solar cycle \citep{KK14,Mea15} but it happened close to the cycle maximum and influenced the polar field dynamics only slightly.
%{\bf
%\citet{JCS15} explained the relatively weak polar fields in the previous %minimum of solar activity by violation of Joy's law at the end of the 23rd %activity cycle.
%}
%%%%%%%%%%%%%%%%%%%%%%%%%%%%%%%%%%%%%%%%%%%%%%%%%%%%%%%%%%%%%%%%%%%%%%%
\section*{Acknowledgements}
This work was supported by the Russian Foundation for Basic Research (project 16--02--00090). The authors are thankful to an anonymous referee for pertinent and constructive comments.
%%%%%%%%%%%%%%%%%%%%%%%%%%%%%%%%%%%%%%%%%%%%%%%%%%%%%%%%%%%%%%%%%%%%%%%
\bibliographystyle{mnras}
\bibliography{kitbib}
%%%%%%%%%%%%%%%%%%%%%%%%%%%%%%%%%%%%%%%%%%%%%%%%%%%%%%%%%%%%%%%%%%%%%%%
% Don't change these lines
\bsp	% typesetting comment
\label{lastpage}
\end{document}